\documentclass[aps,prl,twocolumn,showpacs,preprintnumbers]{revtex4}
\usepackage{graphicx}
\usepackage{amssymb}
\usepackage{amsmath}
\usepackage{xcolor}
\usepackage{times}
\usepackage{latexsym}
\newcommand{\nc}{\newcommand}
\def\beq{\begin{equation}}
\def\eeq{\end{equation}}
\def\bey{\begin{eqnarray}}
\def\eey{\end{eqnarray}}
\nc{\barray}{\begin{eqnarray}}
\nc{\earray}{\end{eqnarray}}
\nc{\barrayn}{\begin{eqnarray*}}
\nc{\earrayn}{\end{eqnarray*}}

\nc{\Tr}{\mbox{Tr}}
\nc{\hc}{\mbox{H.c.}}
\nc{\mptnovx}{p\!\!/_{T,x}}
\nc{\mptnovy}{p\!\!/_{T,y}}
\nc{\spacer}{\phantom{spacer}}

\def\lsim{\mathrel{\raise.3ex\hbox{$<$\kern-.75em\lower1ex\hbox{$\sim$}}}}
\def\gsim{\mathrel{\raise.3ex\hbox{$>$\kern-.75em\lower1ex\hbox{$\sim$}}}}

\nc{\ev}{\,\mathrm{eV}}
\nc{\mev}{\,\mathrm{MeV}}
\nc{\gev}{\,\mathrm{GeV}}
\nc{\tev}{\,\mathrm{TeV}}

\nc{\ttbar}{t\bar t}
\def\sb{\tilde{b}}

\def\st{\tilde{t}}

\nc{\order}[1]{{\mathcal O}(#1)}
\newcommand{\met}{E\!\!\!/_T}
\newcommand{\mpt}{\vec p\!\!\!/_T}
\nc{\mc}{\mathcal}

\begin{document}

\preprint{LA-UR-12-26897}

\title{Hunting Asymmetric Stops}
\author{ Michael L. Graesser$^ a$  and Jessie Shelton$^ b$ }
\affiliation{$^a$Theory Division T-2, Los Alamos National Laboratory, Los Alamos, NM 87545, USA, \\
$^b$  Department of Physics, Harvard University, Cambridge, MA 02138, USA}


\begin{abstract}

  We point out that in the irreducible natural SUSY spectrum, stops
  have comparable branching fractions to chargino-bottom and
  neutralino-top in the vast bulk of parameter space, provided only
  that both decay modes are kinematically accessible.  The total stop
  pair branching fractions into $\ttbar +$ MET can therefore be
  reduced to $\order{50\%}$, while $b\bar b+ X$ branching fractions
  are typically much smaller, $\order{10\%}$, thus limiting the reach
  of traditional stop searches.  We propose a new stop search
  targeting the asymmetric final state $\st\st^*\to t \chi^0 \bar b
  \chi^-+ \hc$, which can restore sensitivity to natural stops in the 7 and 8
  TeV LHC runs.  In addition we present a new variable, {\it topness},
  which efficiently suppresses the dominant top backgrounds to
  semi-leptonic top partner searches.  We demonstrate the utility of
  topness in both our asymmetric search channel and traditional
  $\st\st^*\to\ttbar + $MET searches and show that it matches or
  out-performs existing variables.

\end{abstract}

\maketitle

\section{Introduction}

The recent 
discovery of what looks 
very much 
like a weakly-coupled
Higgs boson has made the question of the electroweak hierarchy acute.
Weak-scale SUSY has long been a favored contender to protect the
Higgs vacuum expectation value, but the impressive agreement of LHC
data with SM predictions has made it clear that most of the MSSM
particle content must lie beyond the kinematic reach of the 7 and 8
TeV LHC if $R$-parity is conserved.  Only a small subset of the SM
superpartners are immediately relevant for the naturalness of the
Higgs potential, however \cite{Cohen:1996vb}.  In particular, only
Higgsinos, with masses $m_H\lsim 200$ GeV, and third-generation
squarks $\st_L,\st_R,\sb_L$, with masses $m_{\tilde Q}\lsim 400$-500 GeV,
need to be light to keep EWSB fully natural, though gluinos, winos,
and binos must not be too far beyond current LHC energies
\cite{Brust:2011tb}.  Searching for a light stop, or three light
third-generation squarks, therefore becomes critical for understanding
the extent to which SUSY is relevant for stabilizing the electroweak
scale.

The purpose of the present work is twofold.  First, we point out that,
in a natural SUSY spectrum, the pure Higgsino nature of the light
neutralinos and chargino can significantly weaken the reach in the
traditional stop search channel, $\st\st^*\to t\bar t + \met$, and we
propose a novel search channel targeting the natural SUSY spectrum
which can recover the lost sensitivity.  Second, we introduce a new
kinematic variable, {\it topness}, which efficiently suppresses the
dominant backgrounds to top partner searches in semileptonic final
states.  We demonstrate its performance in both our novel search
topology and in more traditional searches.

\section{Stop and sbottom branching ratios}

In the natural SUSY spectrum, the lightest states are Higgsinos and
third generation squarks; the gauginos can be heavier, up to several
hundreds of GeV.  Mass mixing between the higgsinos and the heavier EW
gauginos splits the Higgsinos; for example, with
$(M_1,M_2,\mu,\tan\beta)=(400\gev, 800\gev, 200\gev,20)$, the
resulting Higgsino masses are $(m_{\chi_1^0}, m_{\chi^\pm},
m_{\chi_2^0})=(192\gev,197\gev,204\gev)$.

\begin{figure}
\begin{center}
\includegraphics[width=1.1\linewidth]{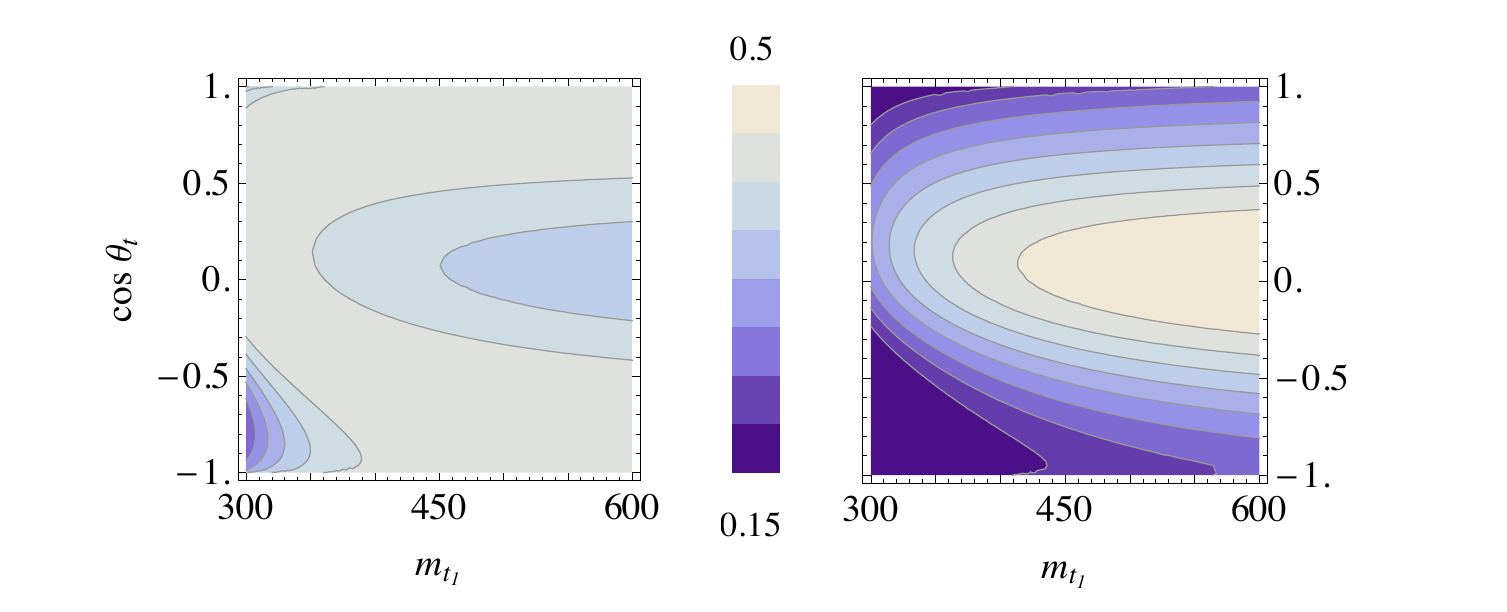}
  \caption{Stop pair branching fractions into (left) the mixed mode
    $t+\chi^0, b+\chi^\pm$; (right) $\ttbar+2\chi^0$, for $\tan \beta =
    20$ and degenerate Higgsinos with mass $\mu = 100$
    GeV.  \label{fig:BRs}}
\end{center}
\end{figure}

The total branching ratios for stop decay into the neutral channel,
$\st_i\to t\chi^0_{1,2}$ and the charged channel, $\st_i\to
b\chi^+_1$, depend only on MSSM Yukawa couplings and phase
space. Provided the mass splitting between stops and higgsinos is
large enough that the top-higgsino mode is open, the stop branching
ratios into charged and neutral higgsino modes are comparable in
magnitude over the large majority of parameter space
\cite{Bartl:1994bu, Baer:2012uy}.  This leads to a sizeable pair-wise
branching ratio into the {\it mixed} final state, $\st \st^* \to t b
\chi^0 \chi^\pm$.  Fig.~\ref{fig:BRs} shows stop pair BRs to both the
mixed channel and the usual search channel $\ttbar+2\chi^0$ in the
$m_{\st}$-$\cos\theta_t$ plane, in the limit
$m_{\chi^0}=m_{\chi^\pm}$.  The total $\sigma\times$ BR into the
$\ttbar +\met$ final state can easily be suppressed by a factor of
two, while the $\sigma\times$ BR into the mixed mode is often
comparable or larger.  The sbottom-like signal, $\st\st^* \to b\bar b
+ 2\chi^\pm$, makes up a distant third, except in regions where the
neutral decay channel is suppressed by available phase-space.

Assuming the gluino is inaccessible, the most stringent collider
bounds on this minimal natural SUSY spectrum are the LEP chargino
bounds \cite{lep}.  We assume for simplicity that the nondegenerate
LEP bound $m_{\chi^\pm}>103.5$ GeV applies, though for sufficiently
small splittings between the chargino and the (N)LSP the limits could
be mildly relaxed.  Limits from direct stop searches
\cite{ATLASstop2,CMSstop} have some reach in the region of interest
($m_{\chi^0} > 100$ GeV, $m_{\st}>m_{\chi^0} +m_t$) but still leave
most of the region unconstrained, particularly when the signal is
reduced by a branching ratio $BR(\st \st\to \ttbar+\met)\sim 0.5$.  In
principle, the CMS razor analysis \cite{CMSrazor} could also place
interesting limits if analysed subject to this signal hypothesis.

In what follows, we use a reference Higgsino sector where the mass
splitting is taken to be $(m_{\chi_1^0}, m_{\chi^\pm},
m_{\chi_2^0})=(m_{\chi_1^0},m_{\chi_1^0}+6\gev,m_{\chi_1^0}+12\gev)$.
This splitting is large enough to reflect the possible influence of
nearby gauginos but small enough that all higgsinos produced in the
decay of 300-600 GeV stops appear primarily as missing energy: the
additional soft daughters produced in the decay of heavier higgsinos
down to the (N)LSP $\chi^0_1$ are not boosted enough to pass selection
cuts for hard isolated objects.  We use a reference stop pair
branching fraction of BR$(\st \st^*\to t b + \met+X)=0.5$, and show
results for a reference signal point with $m_{\st}=500 $ GeV,
$m_{\chi_1^0}=200$ GeV, $\chi^0_1 = \widetilde{h}^0_u$ unless otherwise
specified.  In all signal points we take $\st_1 = \cos \theta_t
\st_R+\sin\theta_t \st_L$ with mixing angle $\cos\theta_t = 0.833$.

We have checked that the modes $\st\to t\chi^0_1$ and $\st\to
t\chi^0_2\to t\chi^0_1+X$ pass our selection cuts with efficiencies
differing by $\mc{O}(10\%)$, with the single largest difference in
efficiency coming from isolated lepton acceptance.  Thus to good
approximation the asymmetric signal is insensitive to the details of
the neutralino mixings.  We have chosen a relatively large mass
splitting for our reference Higgsino sector; in more degenerate
spectra, the difference between $\chi^0_1$ and $\chi^0_2$ would be
negligibly small.

\section{Topness}

The dominant backgrounds to semileptonic stop searches are dileptonic
$\ttbar$ where one of the leptons is either too soft or too forward to
be identified, and $\ttbar$ events with one lepton and one
unidentified $\tau$.  Although three of the six partons in these final
states are missing or unidentified, these backgrounds still contain a
large amount of kinematic information which can be used to identify
events consistent with top quark pair production.

Much literature has been devoted to kinematic variables which can
identify particle masses in the presence of multiple invisible
particles. Two of the most studied variables are the stransverse mass,
$M_{T2}$ \cite{Lester:1999tx}, and the contransverse mass $M_{CT}$
\cite{Polesello:2009rn}. Both of these variables admit straightforward
extensions to the asymmetric decay chains that appear in top
backgrounds with missing leptons, as was studied for $M_{T2}$ in
\cite{Bai:2012gs}.

We propose here a novel alternative.  Dileptonic top events are
reconstructible when both leptons are identified: the mass shell
conditions provide enough constraints to completely solve for the
unmeasured components of the neutrino momenta, up to discrete
combinatoric and quadratic ambiguities.  Once one of the leptons is
lost, this is no longer true: the missing particles are (by
assumption) now a neutrino and a $W$, and one of the mass-shell
conditions is lost along with the lepton, leaving an under-constrained
system.

\begin{figure}
\includegraphics[height=1.1in]{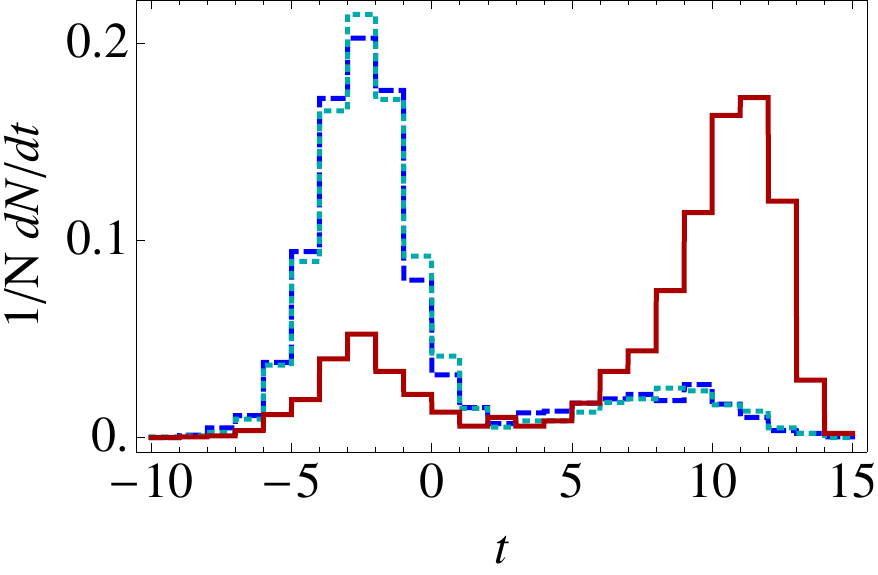} 
\includegraphics[height=1.1in]{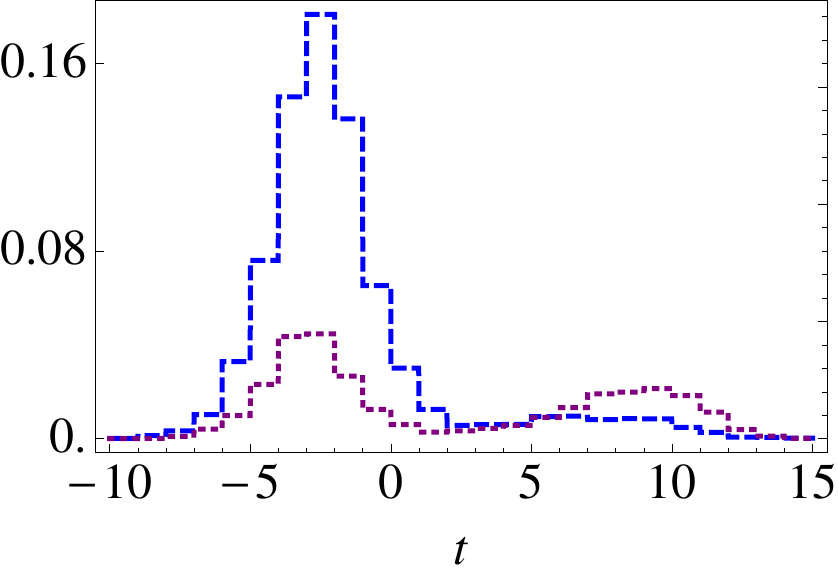}
\caption{Left: unit-normalized topness distributions for events
  passing preselection cuts as described in the text (signal, red,
  solid; dileptonic top, blue, dashed; one $\ell$, one $\tau$ top,
  cyan, dotted). Right: topness distributions for the dileptonic
  background broken down into samples with two truth $b$ jets (blue,
  dashed) and one truth $b$ jet (purple,
  dotted).  \label{fig:topness}}
\end{figure}

\begin{figure}
\includegraphics[height=1.1in]{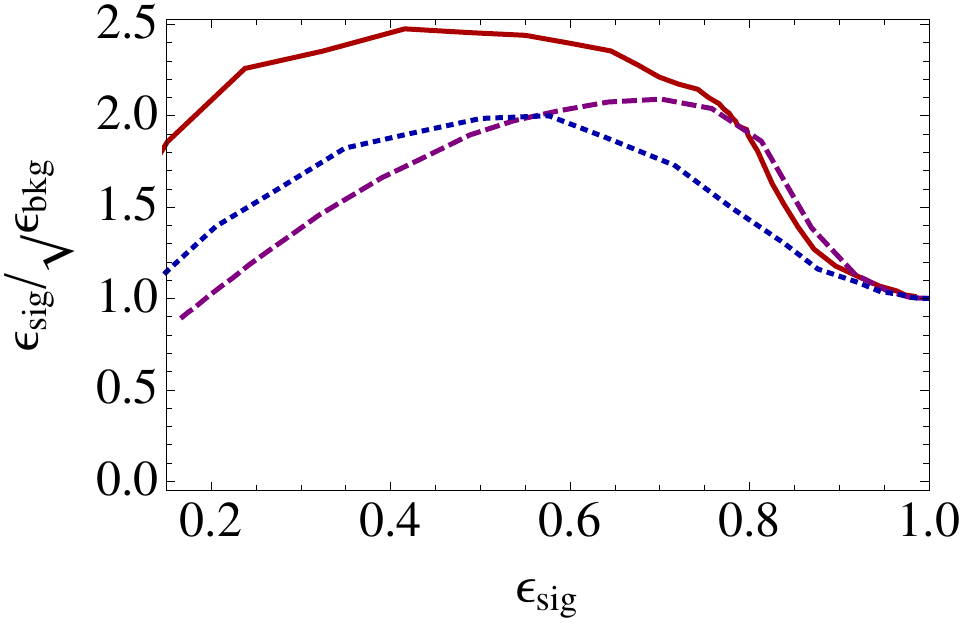} \hspace{2mm}
\includegraphics[height=1.1in]{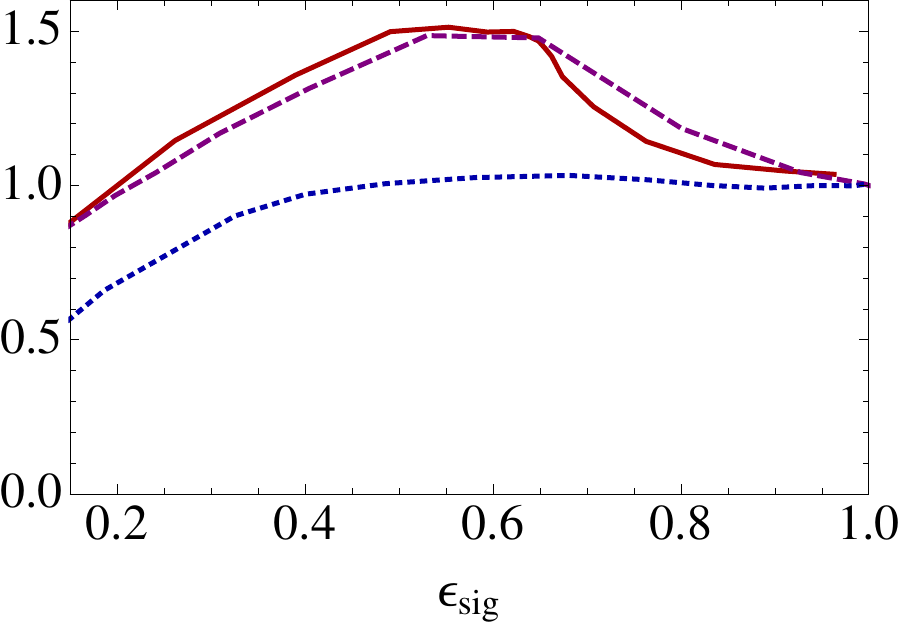}
\caption{  Performance comparison of topness (red, solid) to the
  variables $M_{CT}$ (blue, dotted) and $M_{T2}^W$ (purple, dashed)
  for the asymmetric signal (left) and the symmetric signal 
  $t\bar t + 2\chi^0_1$ (right). The quantity plotted is the gain 
  in signal significance, assuming Gaussian statistics, as a function 
  of signal efficiency.  
  Results are shown for $m_{\st}=500$ GeV and $m_{\chi^0_1}=200$ GeV
  in both channels.
  \label{fig:vsMT2}}
\end{figure}

We replace the missing mass-shell condition with the condition that
{\it the reconstructed center-of-mass energy of the event be
  minimized}. As the PDFs fall off steeply with $\sqrt {s}$, this
provides a good approximation to the true event kinematics.  We
construct a function $S $ which quantifies how well an event can be
reconstructed subject to the dileptonic top hypothesis:
\begin{multline}
S(p_{Wx}, p_{Wy}, p_{Wz}, p_{\nu z}) = \frac{(m_W ^ 2-p_W^2)^2}{a_W^4}  \\
 \:\:+ \frac{(m_t ^ 2-(p_{b_1}+p_\ell+p_\nu)^2)^2}{a_t^4} 
     + \frac{(m_t ^ 2-(p_{b_2}+p_W)^2)^2}{a_t^4}\\
      + \frac{(4m_t ^ 2-(\sum_i p_i)^2)^2}{a_{CM}^4},
  \label{eq:S}
\end{multline}
where in the last term the sum runs over all 5 assumed final state
particles. We have imposed transverse momentum conservation as well as
the mass shell conditions $p_\nu^2 = 0$, $p_W ^ 2 = m_W^2$ to fix
$E_W$, $E_\nu$, $p_{\nu x}$, and $p_{\nu y}$ in terms of the four
remaining undetermined variables.  The denominators $a_k$ determine the
relative weighting of the mass shell conditions, and should not be
smaller than typical resolutions; we take $a_W = 5$ GeV, $a_t = 15$
GeV, and $a_{CM} = 1$ TeV.  The value of $S$ at its minimum quantifies
how well an event can be reconstructed according to the dileptonic top
pair hypothesis.  The inputs to $S$ are two jets, a lepton, and the
$\mpt$.  To find the best possible reconstruction, we sum over both
possible pairings of jets with reconstructed $W $ bosons and keep the
pairing which minimizes $\min S$.  When the event contains two
identified $b$-jets, we use them as input to $S$; when the event
contains only one identified $b$, we consider the two hardest untagged
jets with $|\eta|<2.5$ and $p_T > 20$ GeV, and use the pair $(b,j)$ which
yields the minimum value for $\min S$.  We define {\it topness} as
\beq
t = \ln (\min S) .
\eeq
Minimization of $S$ is a nontrivial computational problem. In our
implementation we use 10 iterations of the Nelder-Mead algorithm per
event. In general this is not sufficient to find the global minimum;
however, it will find a minimum that is sufficiently close to the
global minimum that cuts and distributions are insensitive to any
difference.  We show distributions of topness for the major dileptonic
and one-$\ell$-one-$\tau$ top backgrounds as well as signal in
Fig.~\ref{fig:topness}.  In the left panel of Fig.~\ref{fig:vsMT2} we
compare the performance of topness to both the asymmetric
implementation of $M_{CT}$ which we find most effective, and
$M_{T2}^{W}$, the $M_{T2}$ variant identified as most effective in
\cite{Bai:2012gs}.  Events shown here have passed preselection cuts as
described in the text below.

Our present interest is in the asymmetric stop decay mode, but we
emphasize that topness is useful in any search where the background is
dominated by dileptonic tops with a missed lepton, most notably stop
searches in other channels.  The right panel of Fig.~\ref{fig:vsMT2}
shows the relative performance of topness on the traditional stop
signal, $\st\st ^*\to t\bar t+2\chi^ 0_1$, followed by semileptonic
top decay, with backgrounds given by missed-lepton $t\bar t +2j$
events.  Events shown here have passed object selection cuts analogous
to those in \cite{ATLASstop1}, including the requirement that
$m_T(\ell,\met)>150$ GeV.  Topness performs comparably to $M_{T2}^W$
in the symmetric search channel, realizing (by a slim margin) the
largest gain in significance among all three variables.  The dilution
in the efficacy of topness in the $t\bar t + 2 \chi^0$ channel,
relative to the $tb + \met$ channel, is because the extra jets in the
events give more possibilities for signal to accidentally reconstruct
as a top-like event.  We do comment, however, that our implementation
of topness was not optimized for the $t\bar t + \met$ signal.

\section {Hunting Asymmetric Stops}

In this section we discuss a search strategy for stop pair production
in the mode $\st\st\to t b + $ MET.  We target semileptonic top decay,
so the final state of interest contains one $\ell$, 2 $b$-jets, and
missing energy.  We impose the following cuts at {\it preselection}:
exactly one lepton satisfying $p_\mu>20 $ GeV, $p_e > 25$ GeV,
$|\eta_\ell | < $ 2.5; $\met > 200$ GeV, with $m_T(\ell, \mpt) > 150$
GeV; and at least two jets with $p_T > 20$ GeV and $|\eta|<2.5$, at
least one of which must be $b$-tagged.  The cut on $m_T$ suppresses
all backgrounds where the $\met$ arises from a single $W$, in
particular semileptonic $\ttbar$ and the enormous $W+$ jets, which is
further suppressed by the $b$-tag requirement.  The major remaining
background is therefore dileptonic top pair events where one of the
leptons is not identified, either because it falls outside acceptance,
or because it is a non-identified $\tau$.  A secondary background is
the associated production of a top with a $W$ boson, again with doubly
dileptonic decays and a missed lepton.  All major SM backgrounds can
be reduced by identifying softer leptons in the event; we thus reject
events containing identified hadronic taus with $|\eta|<2.5$ and
$p_T>20$ GeV or additional (isolated) leptons with $|\eta|<2.5$ and
$p_T > 15$ GeV.  Importantly, the additional soft decay products of
the heavier Higgsinos in signal events have negligible impact on the
ability of signal to pass the veto.  More aggressive vetos, as in
\cite{ATLASstop2}, would further reduce the backgrounds at little cost
to signal.

\begin{figure}
\includegraphics[height=1.05in]{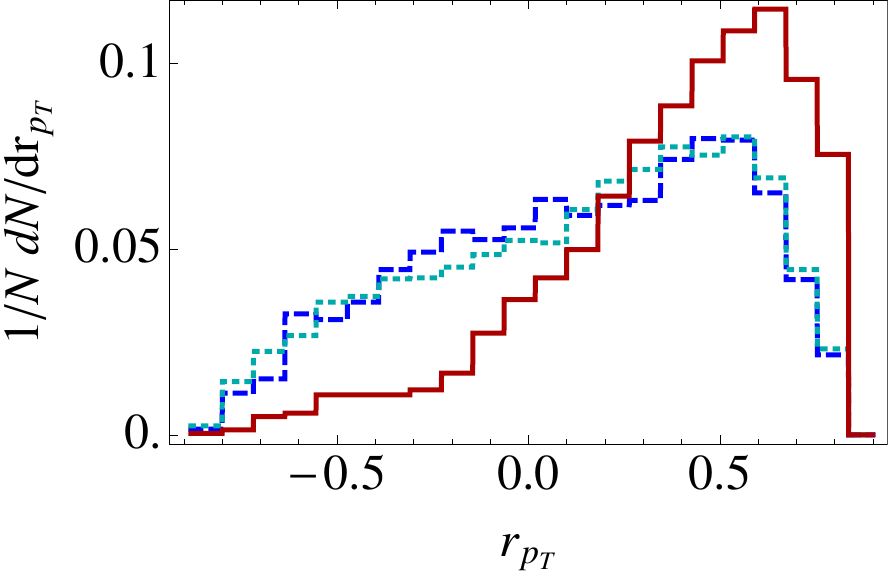}
\includegraphics[height=1.05in]{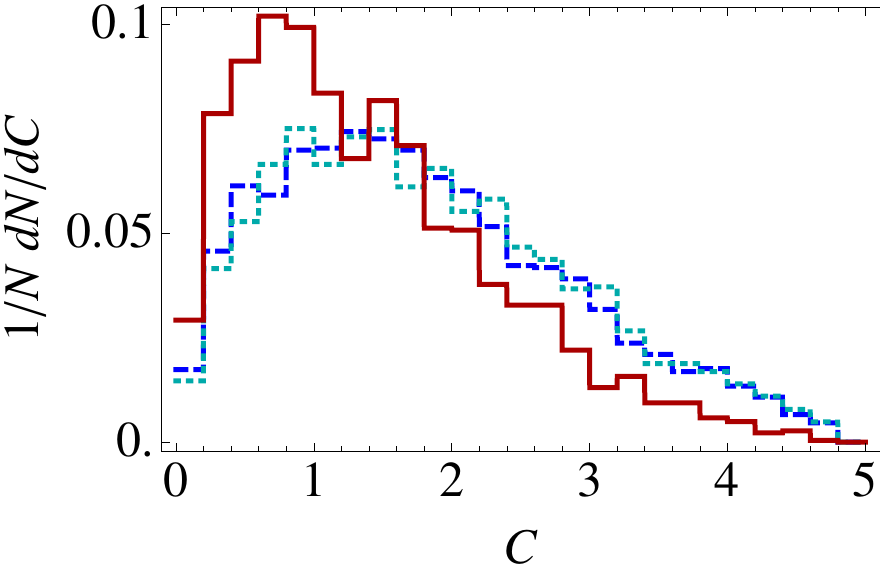}
\caption{Unit-normalized distributions of the variables $r_{p_T}$
  (left) and $C$ (right), for events passing preselection cuts as
  described in the text.  Signal is red and
  solid; dileptonic top is blue and dashed; one $\ell$, one $\tau$ top is
  cyan and dotted.} \label{fig:pTrat}
\end{figure}

In addition to cuts on the hardness of final state particles and
$\met$, we add three novel variables which discriminate signal and
background.  First and by far the most important is topness, discussed
in the previous section.  Another useful variable can be constructed
by exploiting the asymmetric signal kinematics.  The $b$-jet coming
from the decay $\st\to b \chi^+$ is typically much harder than the
daughters of the top.  The $p_T$ asymmetry
\beq
r_{p_T} = \frac{p_{Tb_1}-p_{T\ell}}{p_{Tb_1}+p_{T\ell}}
\eeq
of the lepton and the highest $p_T$ $b$-jet, is thus useful for
distinguishing signal and background.  We also employ a {\it
  centrality} variable,
\beq
C = \max \left(\left|\Delta \eta_{j_1, j_2+\ell+\mpt}\right|, 
                \left|\Delta \eta_{j_2, j_1+\ell+\mpt}\right|\right) ,
\eeq
formed from the two highest $p_T$ jets $j_1$ and $j_2$ in the event as
well as the lepton and missing momentum.  Centrality is typically
larger for backgrounds than for signal, reflecting both the larger
signal masses and the different kinematics of top versus stop pair
production \cite{Han:2012fw}.  Distributions of $r_{p_T}$ and $C$ are
shown in Fig.~\ref{fig:pTrat}.  This particular cut on rapidity
separations is most useful when used in conjunction with topness,
because background events with large topness have often selected an
ISR jet in place of one of the $b$ jets, and this ISR jet is distinct
in rapidity from the other objects in the event.

We normalize signal \cite{Falgari:2012hx} and background $t\bar t $
\cite{Beneke:2012gn} and $tW$ \cite{Kidonakis:2012rm} processes to
inclusive NLO$+$N(N)LL 8 TeV cross-sections.  Events are generated in
Madgraph \cite{Alwall:2011uj}, showered in Pythia
\cite{Sjostrand:2006za}, and clustered in FastJet using the anti-$k_T$
algorithm with $R = 0.4$ \cite{Cacciari:2011ma}.  In generating
$tW+1j$ events, we forbid $t\bar t$ events from contributing when the
momentum in one of the internal top propagators lies in the window
$|p^2-m_t| < 15 \Gamma_t$ \cite{Frixione:2008yi, Alwall:2008qv}.
Leptons are declared isolated if the scalar sum-$p_T$ deposited in a
cone of radius $R_{iso}=0.2$ around a lepton is less than
$r_{iso}=0.2$ times the lepton $p_T$.  The isolation threshold is thus
4 GeV for a lepton with $p_T=20$ GeV, which matches well with
experimentally employed criteria at threshold. 
We consider a hadronic $\tau$ to be identifiable when the scalar sum $p_T$
deposited in a cone of radius $R_{iso} = 0.4$ around the visible
hadronic $\tau$ is less than $r_{iso}= 0.3$ times the $\tau$ $p_T$. We then
apply a probabilistic tagging algorithm to identifiable $\tau$’s to
arrive at a 50\% overall efficiency for identifying hadronic
$\tau$’s with $p_T> 20$ GeV coming from the $W$ in $t\bar t$ events,
comparable to experimental working points.
Finally, we assume a uniform probability
$\mc{P}_b=0.7$ to tag a $b$-jet with $p_T > 20$ GeV and $|\eta|<2.5$.
We then generate more than enough events to ensure
our final results are insensitive to statistical fluctuations in our
Monte Carlo simulations.

\begingroup
\squeezetable
\begin{table}
\begin{center}
\begin{tabular}{l|ccccc}
\hline \hline
                        & $\sigma_{sig}$ &  $\sigma_{t\bar t}$ &  $\sigma_{tW}$ & $S/B $  & $\sigma$ \\
\hline
preselection    &  2.1    & 57   & 5.3   & 0.034 & 1.2     \\

lepton veto     &  2.1 & 43 & 3.9 & 0.045  & 1.4  \\


$b_1$ $p_T>125$ & 1.5 & 21 & 1.6 & 0.065 & 1.4 \\

$r_{pT}>-0.1$     & 1.4 & 20 & 1.5 & 0.067 & 1.4 \\

$C < 4.0$      & 1.4 & 19 & 1.4 & 0.069 & 1.4  \\

$t>9.0$       & 0.89 & 0.62 & 0.37 & 0.90 & 3.5 \\

\hline \hline
\end{tabular}
\caption{Cut flow for the reference signal point, with $m_{\st} = 500$ GeV, 
    $m_\chi = 200 $ GeV, and assuming a 50\% branching ratio into the $tb+\met$ final state.
    All cross-sections are measured in fb, and significances are 
    shown for ${\mathcal L} =  20\:\rm{fb} ^ {-1} $ at 8 TeV.
\label{table:cuteffs}
}
\end{center}

\end{table}

In Table~\ref{table:cuteffs} we show the detailed cut flow for our
reference signal working point as further specialized cuts are
imposed.  As the total number of expected signal and background events
is small, we use Poisson statistics to evaluate signal significance.
Table~\ref{table:sigs} shows optimal cuts for other signal mass
points.  We find in general that the region of best significance is
relatively flat as a function of the cuts, that is, a broad range of
cut combinations yield similar final significances.

\begingroup
\begin{table}
\begin{center}
\begin{tabular}{cc|cccccccc}
\hline \hline
   $m_{\st}$ & $m_{\chi_1^0}$  &  MET & $b_1\,p_T$ & $C$&  $r_{p_T}$ & Topness & $\sigma$ & $S/B$ & $\sigma_{sig}$ (fb) \\
\hline
400 & 100 & 200 & 125 & 3.5 &  -0.1 & 8.0 & 11 & 2.2 & 2.7 \\
400 & 200 & 150 & 50  & 3.5 & -0.1 & 7.5 & 2.7 & 0.31 & 1.4 \\
500 & 100 & 200 & 200 & 4.0 & -0.1 & 9.0 & 4.9 & 1.6 & 1.1 \\
500 & 200 & 200 & 200 & 4.0 & -0.1 & 9.0 & 3.5 & 0.90 & 0.89 \\
600 & 100 & 300 & 300 & 3.5 & -0.3 & 10.5 & 2.9 & 2.2 & 0.32 \\
600 & 200 & 250 & 250 & 4.0 & -0.1 & 10.0 & 2.3 & 1.2 & 0.30 \\
\hline \hline
\end{tabular}
\caption{Signal significances and best cuts for ${\mathcal L} =
  20\:\rm{fb} ^ {-1} $ at 8 TeV.  All masses and energies are in GeV.
  For the signal point ($m_{\st}=400,\, m_{\chi^0_1}=100$), the quoted
  significance is Gaussian.
\label{table:sigs}
}
\end{center}

\end{table}

To conclude, stops in natural SUSY spectra will generically have large
branching fractions to the mixed final state $\st\st^*\to t b + \met +
\mathrm{soft} $, substantially diluting the rate into the traditional
search channel $\st\st\to t\bar t + \met$.  Despite the large $t\bar
t$ backgrounds in the $tb+\met$ final state, good search reach is
achievable with the use of the novel variables introduced here, 
recovering LHC sensitivity to the natural region of parameter space.
The novel variable {\it topness} is critical to this search and is
useful in more generic contexts, matching or outperforming such
kinematic variables as $M_{T2}^W$ and $M_{CT}$.

{\em Acknowledgements:} We thank J.~Gallicchio, T.~Golling, and
M.~Peskin for discussions. MG acknowledges support by the DOE Office
of Science and the LDRD program at Los Alamos National Laboratory.  JS
is supported by the DOE grant DE-FG02-92ER40704, NSF grant
PHY-1067976, and the LHC Theory Initiative.  under grant
NSF-PHY-0969510.  JS thanks the Aspen Center for Physics, under grant
NSF-PHY-1066293, as well as the Galileo Galilei Institute and the
INFN, for hospitality and partial support during the completion of
this work.

\end{document}